\documentclass[prb,12pt,preprint,a4paper,groupedaddress]{revtex4-1}

\usepackage{times}
\usepackage{mathptmx}

\usepackage{amsmath,amssymb,latexsym}
\usepackage{bm}
\usepackage{graphicx}

\newcommand{\p}{\partial}

\newcommand{\vek}[1]{\ensuremath{\mathbf{#1}}}
\newcommand{\uvek}[1]{\ensuremath{\wh{\mathbf{#1}}}}

\newcommand{\wh}[1]{\widehat{#1}}
\newcommand{\wt}[1]{\widetilde{#1}}
\newcommand{\ave}[1]{\left<#1\right>}
\newcommand{\abs}[1]{{\left|#1\right|}}

\newcommand{\Aave}{\langle{A}\rangle}
\newcommand{\Arms}{A_\text{rms}}
\newcommand{\taud}{\tau_\text{d}}
\newcommand{\taup}{\tau_\text{p}}

\newcommand{\m}[1]{\left< #1 \right>}

\newcommand{\four}[2]{\mathcal{F}\left[{#1}\right](#2)}
\newcommand{\fourinv}[2]{\mathcal{F}^{-1}\left[{#1}\right](#2)}
\newcommand{\fourT}[2]{\mathcal{F}_T\left[{#1}\right](#2)}
\newcommand{\psd}[2]{\mathcal{S}_{#1}(#2)}

\let\originalleft\left
\let\originalright\right
\renewcommand{\left}{\mathopen{}\mathclose\bgroup\originalleft}
\renewcommand{\right}{\aftergroup\egroup\originalright}

\newcommand{\Eqref}[1]{Eq.~\eqref{#1}}

\newcommand{\Figref}[1]{Fig.~\ref{#1}}
\newcommand{\Figsref}[1]{Figs.~\ref{#1}}
\newcommand{\Secref}[1]{Sec.~\ref{#1}}

\newcommand{\ARFM}{\textit{Annu.\ Rev.\ Fluid Mech.}}

\newcommand{\ANM}{\textit{Appl. Numer. Math}}

\newcommand{\CPL}{\textit{Chin.\ Phys.\ Lett.}}

\newcommand{\EL}{\textit{Europhys..\ Lett.}}
\newcommand{\FDR}{\textit{Fluid Dyn.\ Res.}}
\newcommand{\GAFD}{\textit{Geophys.\ Astrophys.\ Fluid Dyn.}}
\newcommand{\GRL}{\textit{Geopyhs.\ Res.\ Lett.}}

\newcommand{\JGRA}{\textit{J.\ Geophys.\ Res.\ Atmos.}}

\newcommand{\JPSJ}{\textit{J.~Phys.\ Soc.\ Jpn.}}
\newcommand{\NF}{\textit{Nucl.\ Fusion}}

\newcommand{\PD}{\textit{Phys.~D}}
\newcommand{\PEPI}{\textit{Phys.\ Earth Planet.\ Inter.}}
\newcommand{\PF}{\textit{Phys.\ Fluids}}

\newcommand{\PFB}{\textit{Phys.\ Fluids~B}}
\newcommand{\PPCF}{\textit{Plasma Phys.\ Contr.\ Fusion}}

\newcommand{\PHP}{\textit{Phys.\ Plasmas}}
\newcommand{\PLA}{\textit{Phys.\ Lett.~A}}
\newcommand{\PP}{\textit{Phys.\ Plasmas}}

\newcommand{\PRA}{\textit{Phys.\ Rev.~A}}

\newcommand{\PRE}{\textit{Phys.\ Rev.~E}}
\newcommand{\RMP}{\textit{Rev.\ Mod.\ Phys.}}
\newcommand{\RPP}{\textit{Rep.\ Prog.\ Phys.}}
\newcommand{\PS}{\textit{Phys.\ Scripta}}

\newcommand{\BTSJ}{\textit{Bell Sys.\ Tech. J.}}

\newcommand{\SR}{\textit{Sci.\ Rep.}}

\newcommand{\JFM}{\textit{J.~Fluid Mech.}}
\newcommand{\PRL}{\textit{Phys.~Rev.\ Lett.}}

\newcommand{\NYAS}{\textit{Ann.\ New York Acad.\ Sci.}}

\begin{document}

\title{Intermittent fluctuations due to Lorentzian pulses in turbulent thermal convection}

\author{G.~Decristoforo}
\email{gregor.decristoforo@uit.no}
\author{A.~Theodorsen}
\email{audun.theodorsen@uit.no}
\author{O.~E.~Garcia}
\email{odd.erik.garcia@uit.no}
\affiliation{Department of Physics and Technology, UiT The Arctic University of Norway, NO-9037 Troms{\o}, Norway}

\date{\today}

\begin{abstract}
Turbulent motions due to flux-driven thermal convection is investigated by numerical simulations and stochastic modelling. Tilting of convection cells leads to the formation of sheared flows and quasi-periodic relaxation oscillations for the energy integrals far from the threshold for linear instability. The probability density function for the temperature and radial velocity fluctuations in the fluid layer changes from a normal distribution at the onset of turbulence to a distribution with an exponential tail for large fluctuation amplitudes for strongly driven systems. The frequency power spectral density has an exponential shape, which is a signature of deterministic chaos. By use of a novel deconvolution method, this is shown to result from the presence of Lorentzian pulses in the underlying time series, demonstrating that exponential frequency spectra can persist in also in turbulent flow regimes. 
\end{abstract}

\maketitle

\section{Introduction}\label{intro}

Buoyancy-driven motion of a fluid confined between horizontal plates is a cornerstone of fluid mechanics and has many areas of application, including astrophysics,  industry, laboratory fluid dynamics, meteorology, oceanography, and plasma physics. Due to its rich dynamics, the Rayleigh--B{\'e}nard convection model has become a paradigm to investigate pattern formation, nonlinear phenomena and scaling relationships.\cite{busse-rpp,siggia,bodenschatz,kadanoff,ahlers}

For sufficiently strong forcing, oscillating fluid motion and chaotic behavior results. An intrinsic property of deterministic chaos is an exponential frequency power spectral density for the fluctuations. This has been observed in numerous experiments and model simulations of fluids and magnetized plasmas.\cite{atten,farmer,frisch,greenside,farmer-pd,libchaber,brandstater,sigeti-pra,stone,streett,mensour,paul,reynolds,safonov,schenzinger,sigeti-pd,sigeti-pre,broomhead,osprey,franzke} Recently, the exponential spectrum has been attributed to the presence of uncorrelated Lorentzian pulses in the temporal dynamics.\cite{ohtomo,maggs-pre,pace-prl,pace-pop,hornung,mm-prl,maggs-ppcf,vanmilligen,maggs-ppcf2,zhu,gt-pop1,morales,gt-pop2,gt-pop3,morales2} This includes the Lorenz model, which describes chaotic dynamics in Rayleigh--B{\'e}nard convection.\cite{sigeti-pre,sigeti-pd,broomhead,osprey,franzke,ohtomo,maggs-pre}

In two-dimensional thermal convection, it is well known that the convection rolls in a horizontally periodic domain can give rise to spontaneous formation of strong mean flows through a tilting instability.\cite{busse-pd,howard,drake,fdg,hermiz,prat,rucklidge,fitzgerald,pogutse,sugama,beyer,horton,berning,finn,fh,takayama1,garcia-ppcf,bian,garcia-pre,garcia-pst} For strongly driven thermal convection, turbulent states develop where the sheared mean flows transiently suppress the fluctuating motions, resulting in quasi-periodic relaxation oscillations.\cite{finn,fh,takayama1,garcia-ppcf,bian,garcia-pre,garcia-pst,brummell,busse-clever,gbt,christensen1,aurnou,busse,christensen2,grote,morin,tau,teed} Similar relaxation oscillations have also been identified in turbulent plasmas. \cite{takayama2,guzdar,lin1,lin2,malkov,naulin,takeda,manfredi,esel-php,beyer-prl,esel-ps,kleva} This dynamics has been described in terms of a predator-prey system, with a conservative transfer of kinetic energy from the fluctuating to the mean motions and a viscous dissipation of the latter.\cite{garcia-ppcf,garcia-pre,garcia-pst,bian,bian-php,garcia-php,bian-2010,zcd}
The velocity and temperature fluctuations throughout the fluid layer are strongly intermittent with positive skewness and flatness moments. The probability density functions has exponential tails, resembling the state of hard turbulence in Rayleigh--B{\'e}nard convection.\cite{heslot,castaing,sano,deluca,massaioli,werne,julien,shraiman,cioni,niemela}

In this contribution, it is for the first time demonstrated that these properties of irregular fluid motion can be present simultaneously. The fluctuation statistics in a state of turbulent convection is investigated by numerical simulations of a fluid layer driven by a fixed heat flux.\cite{garcia-ppcf,verzicco,johnston,huang} Time-series analysis and stochastic modelling of the temperature field is presented. It is demonstrated that the frequency power spectral density of the fluctuations has an exponential tail. A novel deconvolution algorithm is applied, showing that the temperature signal can be described as a super-position of Lorentzian pulses. Hence, the well-known properties of deterministic chaos can persist even in turbulent flow regimes.

The outline of this paper is as follows. In \Secref{sec:model} we present the model equations and briefly discuss the shear flow generation mechanism. In \Secref{sec:turbulent} basic results from the numerical simulations are presented. The fluctuation statistics are presented in \Secref{sec:fluctuations} and in \Secref{sec:lorentzian} it is demonstrated that the exponential frequency power spectral density is due to the presence of Lorentzian pulses in the time series. The conclusions and a summary of the results are presented in \Secref{sec:concl}. The appendix presents a derivation of the frequency power spectral density due to a periodic train of pulses with fixed shape and duration.

\section{Model equations}\label{sec:model}

Considering two-dimensional fluid motions in a gravitational field opposite to the $x$-axis, the model equations describing thermal convection are given by
\begin{subequations}
\begin{gather}
\left( \frac{\p}{\p t} + \uvek{z}\times\nabla\psi\cdot\nabla \right) \Theta = \kappa\nabla^2\Theta , \label{temperature}
\\
\left( \frac{\p}{\p t} + \uvek{z}\times\nabla\psi\cdot\nabla \right) \Omega + \frac{\p\Theta}{\p y} = \mu\nabla^2\Omega , \label{vorticity}
\end{gather}
\end{subequations}
where $\Theta$ describes the temperature, $\psi$ is the stream function for the two-dimensional fluid velocity field $\vek{v}=\uvek{z}\times\nabla\psi$, and $\Omega=\uvek{z}\cdot\nabla\times\vek{v}=\nabla^2\psi$ is the associated fluid vorticity. The temperature perturbations are normalized by the temperature difference $\triangle T$ over the fluid layer in hydrostatic equilibrium, length scales are normalized by the fluid layer depth $d$, and time is normalized by the ideal interchange rate.\cite{verzicco,johnston,huang} The normalized heat diffusivity $\kappa$ and viscosity $\nu$ are related to the Rayleigh and Prandtl numbers by $R=1/\kappa\nu$ and $P=\nu/\kappa$, respectively. The temperature in hydrostatic equilibrium is given by $\Theta=1-x$. A similar mathematical model also describes fluctuations in non-uniformly magnetized plasmas where the symmetry axis $z$ corresponds to the direction of the magnetic field and effective gravity is due to magnetic field curvature.\cite{pogutse,sugama,beyer,horton,berning,finn,fh,takayama1,garcia-ppcf,bian,garcia-pre,garcia-pst}

In many cases, the fluid is confined in a geometry where $x$ corresponds to the radial coordinate and $y$ the azimuthal direction. All dependent variables are accordingly assumed to be periodic in the $y$-direction, for example $\Theta(y)=\Theta(y+L)$. In the radial direction the boundary conditions are taken to be
\begin{subequations}
\begin{gather}
\psi(x=0) = \psi(x=1) = 0 ,
\\
\Omega(x=0) = \Omega(x=1) = 0 ,
\\
\frac{\p\Theta}{\p x}(x=0) = -1, \quad \Theta(x=1) = 0 .
\end{gather}
\end{subequations}
The latter condition corresponds to a fixed conductive heat flux through the fluid layer.\cite{garcia-ppcf,verzicco,johnston,huang} It should be noted that the free-slip boundary conditions imply that there is no convective heat transport through the radial boundaries since $v_x=-\p\psi/\p y=0$ for $x=0,1$.

For the azimuthally periodic system, it is convenient to define the profile of any dependent variable as its azimuthal average and denote this by a zero subscript. For the temperature field $\Theta$ this is given by
\begin{equation}
\Theta_0(x,t) = \frac{1}{L}\int_0^L \text{d}y\,\Theta(\vek{x},t) .
\end{equation}
The motivation for separating profiles and spatial fluctuations is simply that the latter are the components mediating radial convective heat flux while the former describes modifications of the equilibrium state profiles.

Similar to the temperature profile, an average azimuthal flow is also defined by
\begin{equation}
v_0(x,t) = \frac{1}{L} \int_0^L \text{d}y\,\frac{\p\psi}{\p x} = \frac{\p\psi_0}{\p x} .
\end{equation}
Due to conservation of net circulation of the fluid layer, the mean azimuthal flow is intrinsically sheared and corresponds to differential rotation of the fluid layer. Such flows develop due to a tilting instability of the convective cells.\cite{busse-pd,howard,drake,fdg,hermiz,prat,rucklidge,fitzgerald,pogutse,sugama,beyer,horton,berning,finn,fh,takayama1,garcia-ppcf,bian,garcia-pre,garcia-pst} Since the symmetric flow $v_0$ is intrinsically incapable of mediating radial convective transport, it is natural to separate the kinetic energy into two components comprised by the fluctuating motions and the sheared mean flows, defined respectively by
\begin{equation}
K(t) = \int \text{d}\vek{x}\,\frac{1}{2}\,[\nabla(\psi-\psi_0)]^2 ,
\qquad
U(t) = \int \text{d}\vek{x}\,\frac{1}{2}\,v_0^2 .
\end{equation}
The evolution of these energy integrals are readily derived from the mean vorticity equation,\cite{garcia-ppcf,bian,garcia-pre,garcia-pst}
\begin{align}
\frac{\text{d}K}{\text{d}t} & = \int\text{d}\vek{x}\,v_x\Theta - \Pi - \nu \int\text{d}\vek{x}\,(\Omega-\Omega_0)^2 ,
\\
\frac{\text{d}U}{\text{d}t} & = \Pi - \nu \int\text{d}\vek{x}\,\Omega_0^2 ,
\end{align}
where the kinetic energy transfer rate from from the fluctuating motions to the sheared mean flows is defined by
\begin{equation}
\Pi = \int\text{d}\vek{x}\,v_0\frac{\p}{\p x}({v}_x{v}_y) .
\end{equation}
As expected, the convective transport drive for the kinetic energy integral in equation (21b) appears only for the fluctuating motions, while viscous dissipation damps kinetic energy in either form. Radial convective transport of azimuthal momentum evidently yields a conservative transfer of kinetic energy between the fluctuating motions and the azimuthally mean flows.

Numerical simulations have shown that turbulent convection can display predator-prey-like relaxation oscillations for the energy integrals,\cite{finn,fh,takayama1,garcia-ppcf,garcia-pre,garcia-pst} which can be interpreted as follows. Initially the convective energy grows exponentially due to the primary instability. When the fluctuation level becomes sufficiently large to sustain the sheared mean flows against viscous dissipation, this flow energy grows at the expense of the convective motions. The spatial fluctuations are effectively stabilized at a sufficiently strong shear flow. Kinetic energy is however continuously transferred to the mean flows, leading to an almost complete suppression of the fluctuation energy and thus the radial convective transport. Subsequently, there are no fluctuating motions to sustain the sheared flows, which hence decay on a viscous time scale. Finally, as the mean flows become sufficiently weak, the convective energy again starts to grow and the cycle repeats. As will be seen from the numerical simulations presented in the following section, this leads to a strong modulation of the fluctuations.

\section{Turbulent convection}\label{sec:turbulent}

The temperature and vorticity equations are solved numerically by combining a finite difference and a Fourier-Galerkin method for spatial discretization, using an Arakawa scheme for exact conservation of energy and enstrophy. For time discretization a third order stiffly stable integrator is used.\cite{nn,esel-prl,esel-php} Time series of the dependent variables are recorded at radially equi-distant points in the simulation domain and analyzed in the following.

For sufficiently high Rayleigh numbers, numerical simulations of the two-dimensional thermal convection model results in turbulent states.\cite{garcia-ppcf,garcia-pre,garcia-pst} Previously, it has been shown that close to the onset of turbulent convection, for $R=4\times10^5$ and $P=1$, the radial velocity fluctuations in the center of the domain are normally distributed.\cite{garcia-ppcf} Increasing the Rayleigh number to $R=2\times10^6$ results in a probability distribution function for the radial velocity fluctuations with exponential tails.\cite{garcia-ppcf,garcia-pre,garcia-pst} Previous investigations have shown that the large-amplitude fluctuations are associated with coherent structures propagating through the fluid layer. Here we present a detailed analysis of the fluctuation statistics in latter parameter regime (with $R=2\times10^6$, $P=1$ and $L=1$), resembling the state of hard turbulence in thermal convection experiments.\cite{heslot,castaing,sano,deluca,massaioli,werne,julien,shraiman,cioni,niemela}

\begin{figure}[t]
\centering
\includegraphics[width=8cm]{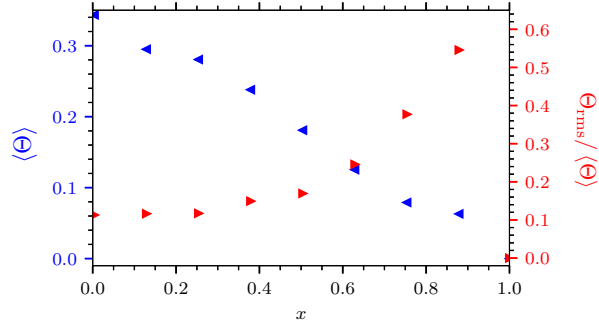}
\caption{Time-averaged profile of the temperature and the relative fluctuation level.}
\label{tprofile}
\end{figure}

The time-averaged profile of the temperature as well as the relative fluctuation level is presented in \Figref{tprofile}. Here and in the following, angular brackets indicate a time average. The turbulent motions significantly reduce the heat confinement in the fluid layer, reducing the temperature on the left boundary from unity in the case of only heat conduction to less than $0.343$ on average in the turbulent state. There is a significant profile gradient in the center of the fluid layer. The relative fluctuation level increases drastically from the center of the domain and radially outward, reaching more than 0.5 close to the outer boundary.

\begin{figure}[t]
\centering
\includegraphics[width=8cm]{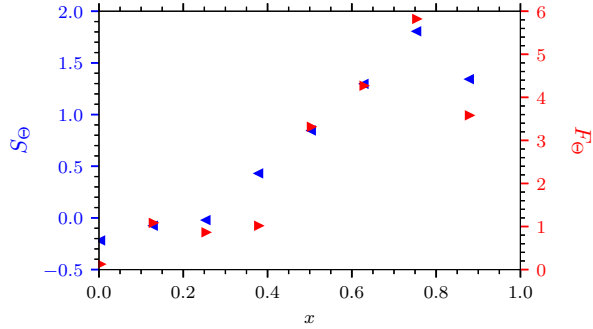}
\caption{Time-averaged profile of the skewness and flatness moments for the temperature fluctuations.}
\label{skprofiles}
\end{figure}

The intermittency of the fluctuations is quantified by the skewness moment, defined by $S_\Theta=\langle(\Theta-\ave{\Theta})^3\rangle/\Theta_\text{rms}^3$, and the flatness moment, defined by $F_\Theta=\langle(\Theta-\ave{\Theta})^4\rangle/\Theta_\text{rms}^4-3$, where the variance is given by $\Theta_\text{rms}^2=\langle(\Theta-\ave{\Theta})^2\rangle$. Both of the skewness and flatness moments vanish for a normally distributed random variable. The profile of these moments for the temperature fluctuations are presented in \Figref{skprofiles}. This shows that the probability density for the fluctuations is positively skewed and flattened in the outer part of the simulation domain, suggesting frequent appearance of large-amplitude bursts in the time series at a fixed point in the fluid layer. The moments are largest at $x=3/4$, where the skewness is 1.81 while the flatness moment is 5.82. This demonstrates a strong departure from a normal distribution of the fluctuations.

\begin{figure}[t]
\centering
\includegraphics[width=8cm]{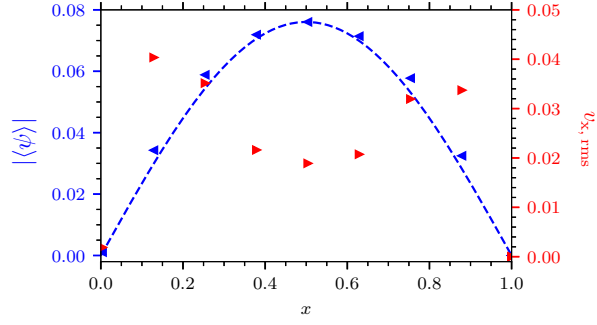}
\caption{Time-averaged profile of the stream function and the root mean square value of the radial velocity. The broken line shows a half-period sine function fit to the stream function profile.}
\label{fig:vprofiles}
\end{figure}

The time-averaged profile of the stream function is presented in \Figref{fig:vprofiles} together with the root mean square fluctuation level of the radial velocity. The time-averaged stream function has a near half-period sinoidal variation over the fluid layer and vanishes at the boundaries. This implies an average counter-streaming mean flow in the fluid layer which vanishes in the center of the domain and is strongest close to the boundaries. However, the radial velocity fluctuation vanishes at the boundaries due to the stress--free boundary conditions. The velocity fluctuation level has a local minimum in the center of the domain. At $x=3/4$ the mean flow is $0.155$, resulting in a vertical transit time of approximately $6.46$ in non-dimensional units. There are some changes in this transit time since the mean flow velocity changes in time, as discussed later.

\begin{figure}[t]
\centering
\includegraphics[width=8cm]{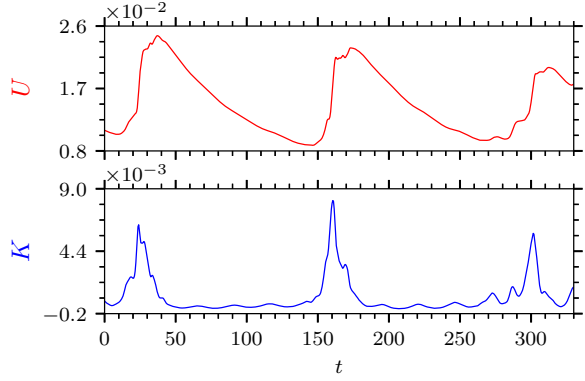}
\caption{Evolution of the kinetic energy in fluctuating, $K$, and mean flows, $U$, showing predator-prey-like relaxation osciallations.}
\label{fig:energy}
\end{figure}

\begin{figure}[t]
\centering
\includegraphics[width=8cm]{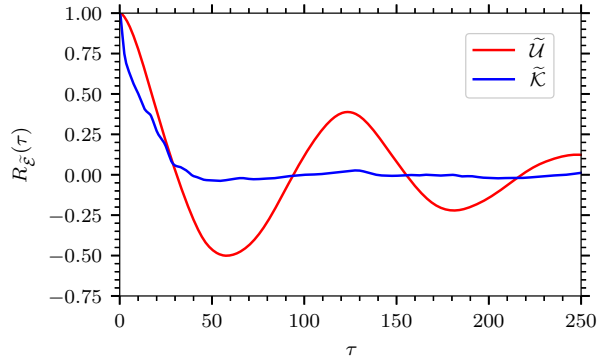}
\caption{Auto-correlation function for the kinetic energy in fluctuating and mean flows.}
\label{fig:energyacf}
\end{figure}

The evolution of the kinetic energy in the fluctuating and mean motions for a short part of the simulation run is presented in \Figref{fig:energy}. This shows the quasi-periodic relaxation oscillations resembling predator--prey type dynamics, where kinetic energy is transferred from the fluctuating motions to the sheared flows and subsequently dissipated by viscosity. The auto-correlation function for the energy integrals are presented in \Figref{fig:energyacf}. The mean flow energy auto-correlation function has a damped oscillatory behavior with period of approximately $125$, corresponding to the characteristic separation between bursts in the energy integrals. The auto-correlation function for the energy in the fluctuating motions has a decay time of approximately $25$, which is attributed to the  characteristic duration of the bursts in the kinetic energy seen in \Figref{fig:energy}. This has been confirmed by conditional averaging of large-amplitude events in the energy integral time series.

\begin{figure}[t]
\centering
\includegraphics[width=8cm]{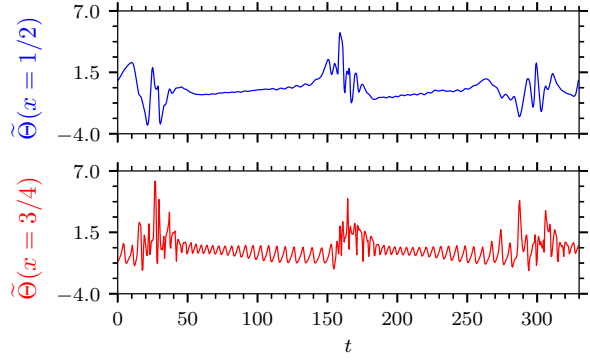}
\caption{Time series of the temperature fluctuations at $x=1/2$ and $x=3/4$.}
\label{fig:traw}
\end{figure}

In \Figref{fig:traw} the normalized temperature fluctuations $\wt{\Theta}=(\Theta-\ave{\Theta})/\Theta_\text{rms}$
recorded in the center of the fluid layer, at $x=1/2$, and in the outer part, at $x=3/4$, are presented. Throughout the fluid layer, the fluctuations are strongly intermittent with large bursts during the time of strong activity in the energy of the fluctuating motions presented in \Figref{fig:energy}. In the outer part of the fluid layer, the fluctuations have a nearly periodic oscillation in the periods between the bursts in the energy integrals. This is due to the sheared mean flow with a transit time of approximately $6.46$. In the following, the statistical properties of these fluctuations will be elucidated.

\section{Fluctuation statistics}\label{sec:fluctuations}

\begin{figure}[t]
\centering
\includegraphics[width=8cm]{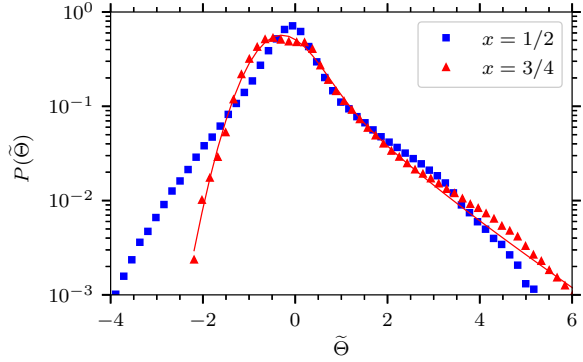}
\caption{Probability density function for the normalized temperature fluctuations at $x=1/2$ and $x=3/4$. The full lines show the best fit of a convolution between a Gamma and a normal distribution.}
\label{fig:tpdf}
\end{figure}

\begin{figure}[t]
\centering
\includegraphics[width=8cm]{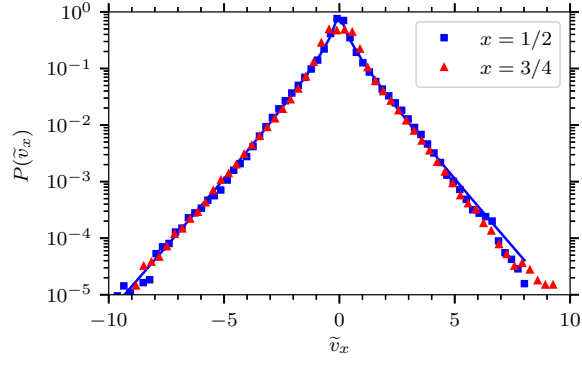}
\caption{Probability density function for the normalized radial velocity fluctuations at $x=1/2$ and $x=3/4$. The full line show the best fit of a convolution between a Laplace and a normal distribution.}
\label{fig:vpdf}
\end{figure}

The probability density function for the temperature fluctuations measured at $x=1/2$ and $x=3/4$ are presented in \Figref{fig:tpdf}. As expected from the radial variation of the skewness and kurtosis moments, the distributions have elevated tails compared to a normal distribution. For $x=3/4$ the distribution is strongly skewed and has a nearly exponential tail towards large values. This is demonstrated by the full line in \Figref{fig:tpdf}, which is the best fit of a convolution of a normal distribution and a Gamma distribution. Similarly, the probability distribution functions for the radial velocity fluctuations are presented in \Figref{fig:vpdf} together with the best fit of a convolution between a Laplace and a normal distribution. This clearly demonstrates the presence of exponential tails in the probability densities.

\begin{figure}[t]
\centering
\includegraphics[width=8cm]{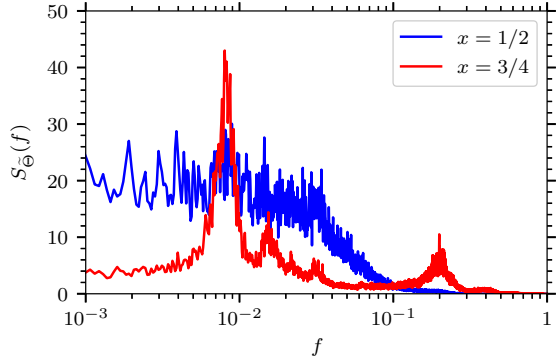}
\caption{Logarithm of the frequency power spectral density for temperature fluctuations at $x=1/2$ and $x=3/4$.}
\label{fig:tpsdlog}
\end{figure}

\begin{figure}[t]
\centering
\includegraphics[width=8cm]{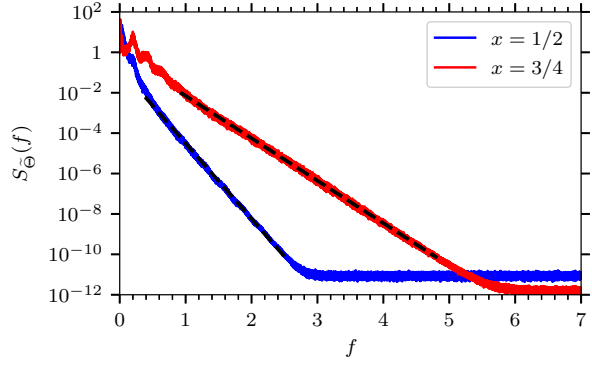}
\caption{Frequency power spectral density for temperature fluctuations at $x=1/2$ and $x=3/4$. The broken lines show the best fit of an exponential function.}
\label{fig:tpsdflog}
\end{figure}

The frequency power spectral densities for the temperature signals measured at $x=1/2$ and $x=3/4$ are presented in semi-logarithmic plots in \Figsref{fig:tpsdlog} and \ref{fig:tpsdflog}. From \Figref{fig:tpsdlog}, with logarithmic scaling of the frequency, it is clear that the frequency spectrum has a pronounced maximum at the linear frequency $f=8\times10^{-3}$, which correspond to the characteristic time between bursts in the energy integrals discussed above. Some higher harmonics of this frequency peak are also readily identified. The frequency spectrum for $x=3/4$ also has a peak at approximately $f=0.2$, corresponding to the vertical transit time by the average mean flow.

When the spectra are plotted with a logarithmic scaling for the power as presented in \Figref{fig:tpsdflog}, it is clearly seen that frequency power spectral density has an exponential decay on the form $\exp(-4\pi \taud\abs{f})$, with the characteristic time $\taud=0.637$ for $x=1/2$ and $\taud=0.382$ for $x=3/4$. In the following section, it will be demonstrated that the exponential spectrum is due to the presence of Lorentzian pulses in the time series and that the slope corresponds to the duration time of these pulses.

\section{Lorentzian pulses}\label{sec:lorentzian}

An exponential frequency power spectrum is a signature of deterministic chaos and has been attributed to Lorentzian-shaped pulses in the underlying time series.\cite{} In order to demonstrate this, consider the stochastic process give a super-position of pulses with fixed shape $\phi$ and duration $\taud$\cite{rice1,daly,garcia-prl,kg-mse,bergsaker,theodorsen-x,gktp,militello,theodorsen-ps,gt,theodorsen-pdf,theodorsen-x2}
\begin{equation} \label{shotnoise}
\Phi_K(t) = \sum_{k=1}^{K(T)} A_k \phi\left( \frac{t-t_k}{\taud} \right) ,
\end{equation}
where $A_k$ and $t_k$ are pulse amplitude and arrival time for the pulse labeled $k$, and $K(T)$ is the number of pulses present in a time interval of duration $T$. In the case of Lorentzian pulses, the function $\phi$ is given by\cite{gt-pop1,gt-pop2,gt-pop3}
\begin{equation} \label{Lorentzian}
\phi(\theta) = \frac{1}{\pi}\frac{1}{1+\theta^2} .
\end{equation} 
In the case of uncorrelated Lorentzian pulses it was recently shown that the frequency power spectral density is exponential, and for the normalized variable $\wt{\Phi}=(\Phi-\ave{\Phi})/\Phi_\text{rms}$ given by\cite{gt-pop1,gt-pop2}
\[
\mathcal{S}_{\wt{\Phi}}(f) = 2\pi\taud\exp( - 4\pi\taud \abs{f} ) .
\]
In the appendix, it is shown that for a periodic sequence of Lorentzian pulses with fixed duration, the frequency power spectrum is a product of the exponential spectrum and a uniform delta pulse train at frequencies corresponding to multiples of the inverse periodicity time. In the case of a slight irregularity in the period between the pulses, the delta peaks in the frequency spectrum with broaden and have finite amplitude.

\begin{figure}
\centering
\includegraphics[width=8cm]{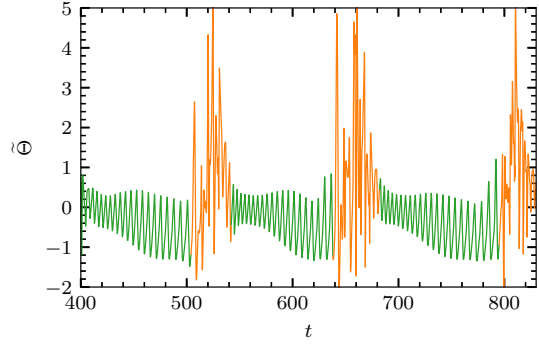}
\caption{Example of splitting of temperature fluctuations into quasi-periodic (green) and turbulent (orange) parts at $x=3/4$.}
\label{fig:turb-per-ex}
\end{figure}

\begin{figure}
\centering
\includegraphics[width=8cm]{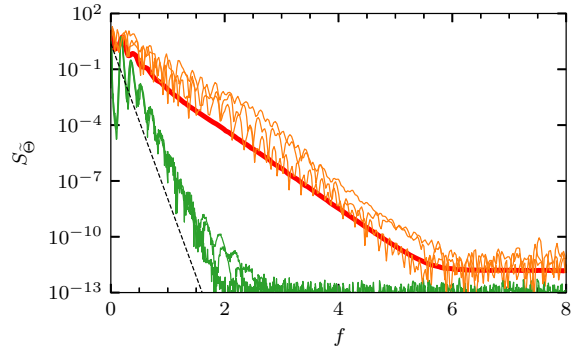}
\caption{Power spectral density of the temperature fluctuations at $x=3/4$. The red line gives the spectrum of the entire time series as shown in \Figref{fig:tpsdlog}, the orange lines give the spectra of the turbulent parts from \Figref{fig:turb-per-ex}, while the green lines give the spectra of the quasi-periodic parts. The black dashed line gives the prediction for an exponential spectrum with the periodicity time seen in the raw time series.}
\label{fig:turb-per-psd}
\end{figure}

As seen in \Figref{fig:traw}, the temperature time series at $x=3/4$ can be separated into parts with nearly periodic oscillations and turbulent parts with chaotic, large-amplitude fluctuations. An example of separating these periods is shown in \Figref{fig:turb-per-ex}. In \Figref{fig:turb-per-psd}, the frequency power spectral density of the quasi-periodic and turbulent parts are shown together with the power spectral density of the entire signal. It is clear that the power spectrum of the entire signal is well described by the power spectrum of the turbulent parts, and that they have the same time scale, $\taud=0.382$. The black dashed line gives an exponential spectrum with 4 times the duration time of the whole spectrum, which closely resembles the power spectrum of the quasi-periodic parts of the signal. In the following, this is used as an estimated duration time of the periodic part.

In order to demonstrate that the temperature time series can be described as a super-position of Lorentzian pulses, a deconvolution algorithm using a Lorentzian pulse with fixed duration time estimated from the power spectral density is applied. This gives the pulse amplitudes and arrival times, which can be used to reconstruct the original signal. The super-position of pulses with fixed duration given by \Eqref{shotnoise} can be written as a convolution between the pulse function and a train of delta pulses,\cite{theodorsen-ps,tg-php}
\begin{equation} \label{eq:fpp-convolve}
\Phi_K(t) = \left[ \phi * F_K \right]\left(\frac{t}{\taud}\right),
\end{equation}
where
\begin{equation} \label{eq:fK}
F_K(t) = \sum\limits_{k=1}^{K(T)} A_k \delta\left( \frac{t - t_k}{\taud} \right).
\end{equation}
The goal is to find the forcing $F_K(t)$ and to estimate the pulse amplitudes $\{A_k\}_{k=1}^K$ and arrival times $\{t_k\}_{k=1}^K$ as accurately as possible. In order to do this, a modified version of the Richardson--Lucy deconvolution algorithm will be used.\cite{tg-php,richardson,lucy,witherspoon,dellacqua,tai} Following this scheme, an initial guess for $F_K$ is made, denoted by $F_K^{(1)}$. The numerical value of this initial forcing matters little, and can be set as some positive constant or the signal itself. The initial value is updated iteratively, with the $n$'th iteration given by
\begin{equation}\label{eq:rl-deconv}
F_K^{(n+1)} = F_K^{(n)} \frac{D*\widehat{\phi}}{F_K^{(n)}*\phi*\widehat{\phi}},
\end{equation}
where $\widehat{\phi}(t)=\phi(-t)$. Here and in the following, $D$ denotes any of the simulation data time series discussed above.

\begin{figure}
\centering
\includegraphics[width=8cm]{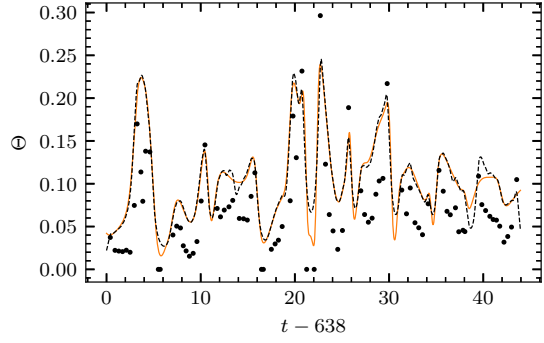}
\caption{Excerpt of turbulent part (full orange line) and reconstructed signal from the deconvolution (black dashed line). The dots indicate arrival times and amplitudes of Lorentzian pulses with duration time $\taud=0.382$. The circular dots give half the true amplitude value for better comparison with the time series.}
\label{fig:turb-rec-1}
\end{figure}

\begin{figure}
\centering
\includegraphics[width=8cm]{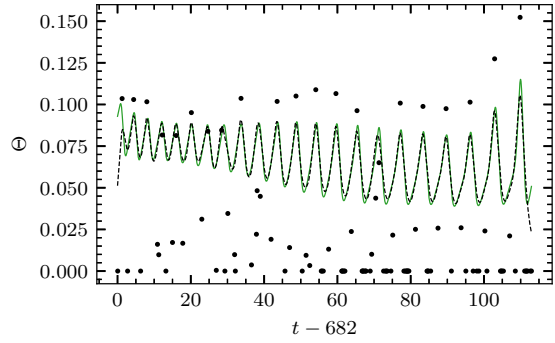}
\caption{Excerpt of quasi-periodic part (full green line) and reconstructed signal from the deconvolution (black dashed line). The dots indicate arrival times and amplitudes of Lorentzian pulses with duration time $\taud=1.528$. The circular dots give half the true amplitude value for better comparison with the time series.}
\label{fig:per-rec-2}
\end{figure}

The result from the deconvolution algorithm is presented in \Figsref{fig:turb-rec-1} and \ref{fig:per-rec-2} for representative turbulent and the quasi-periodic parts, respectively. It is clear that most of the signal is well reconstructed by a super-position of Lorentzian pulses. The frequency power spectral density of the reconstructed time series accurately reproduces that from the numerical simulations as expected. This analysis clearly demonstrates that that the exponential frequency spectra for the temperature fluctuations in the thermal convection simulations are due to the presence of Lorentizan pulses in the time series.

\section{Discussion and conclusions}\label{sec:concl}

In this contribution, the statistical properties of the temperature fluctuations in numerical simulations of turbulent thermal convection have been investigated by time-series analysis and stochastic modelling. The generation of a sheared mean flow through the fluid layer results in predator-prey-like dynamics of the energy integrals and leads to multiple temporal scales in the dynamics. For sufficiently large Rayleigh numbers, a regime corresponding to hard turbulence results with exponential tails in the probability distribution function for the temperature and velocity fluctuations.

The frequency power spectral density for the fluctuations have local maxima at frequencies corresponding to bursting in the energy integral as well as transit time for the mean flow through the fluid layer. However, when presented in a semi-logarithmic plot it is clear that the frequency spectrum has an exponential tail for power densities all the way down to round off errors. A novel deconvolution method has been used to show that the exponential spectrum is due to the presence of Lorentzian pulses in the temperature time series. The time scale for the structures is consistent with the slope of the exponential frequency spectra.

\section*{Acknowledgements}

This work was supported by the UiT Aurora Centre for Nonlinear Dynamics and Complex Systems Modelling. Audun Theodorsen was supported by a Tromsø Science Foundation Starting Grant.

\section*{Data availability}

The data that support the findings of this study are available from the corresponding author upon reasonable request.

\appendix

\section{}

In this appendix, the frequency power spectral density for a super-position of pulses with periodic arrivals is calculated. A super-position of $K$ pulses with fixed shape and duration, as given by \Eqref{shotnoise}, can be written as a convolution between the pulse function $\phi$ and a train of delta pulses,
\begin{equation} \label{eq:shot_noise_green}
\Phi_K(t) = \int_{-\infty}^\infty \text{d}\theta\,\phi\left( \frac{t}{\taud}-\theta \right) F_K(\theta) ,
\end{equation}
where the forcing $F_K$ due to the delta pulse train is given by
\begin{equation}  \label{eq:f}
F_K(\theta) = \sum_{k=1}^K A_k \delta\left(\theta - \frac{t_k}{\taud} \right) .
\end{equation}
The pulse duration time $\taud$ is taken to be the same for all pulses and the pulse amplitudes $A_k$ are taken to be randomly distributed with mean value $\Aave$ and variance $\Arms^2=\langle(A-\Aave)^2\rangle$. The pulse function is assumed to be localized and normalized such that\cite{gktp,gt}
\[
\int_{-\infty}^\infty \text{d}\theta\,\abs{\phi(\theta)} = 1 .
\]

The frequency power spectral density of a random process $\Phi_K(t)$ is defined as
\begin{equation}
\psd{\Phi}{\omega} = \lim_{T \to \infty} \m{\abs{\fourT{\Phi_K}{\omega}}^2} ,
\end{equation}
where the Fourier transform of the random variable over the domain $[0,T]$ is defined by
\begin{equation}
\fourT{\Phi_K}{\omega} = \frac{1}{\sqrt{T}} \int\limits_{0}^{T} \text{d}t\,\exp(-i \omega t) \Phi_K(t) .
\end{equation}
Here $\omega=2\pi f$ is the angular frequency. Analytical functions which fall sufficiently rapid to zero, such as the pulse function $\phi$, have the Fourier transform 
\begin{equation}
\four{\phi}{\vartheta} = \int\limits_{-\infty}^{\infty} \text{d}\theta\,\exp(-i \theta \vartheta)\phi(\theta)
\end{equation}
and the inverse transform
\begin{equation} \label{eq:appendix_fourinv}
\phi(\theta) = \fourinv{\four{\phi}{\vartheta}}{\theta} =  \frac{1}{2\pi} \int\limits_{-\infty}^{\infty} \text{d}\vartheta\,\exp( i \theta \vartheta ) \four{\phi}{\vartheta} .
\end{equation}
Note that here, $\theta$ and $\vartheta$ are non-dimensional variables, as opposed to $t$ and $\omega$.

Neglecting end effects in by assuming $T/\taud\gg1$, the frequency power spectral density of the stationary process $\Phi_K$ is found to be the product of the power of the pulse function and the power of the forcing,\cite{theodorsen-ps}
\begin{equation} \label{eq:psd_sn_start}
\psd{\Phi}{\omega} = \abs{\four{\phi}{\taud \omega}}^2 \lim_{T \to \infty} \m{\abs{\fourT{F_K\left(\frac{t}{\taud}\right)}{\omega}}^2} ,
\end{equation}
which is independent of $K$ since the average is over all random variables. The frequency power spectrum for the forcing $F_K$ will now be calculated for the case of periodic pulses.

The marginal probability density function for the pulse arrival times when these are periodic with period $\taup$ and starting point $s$, assuming the starting time $s$ is known, is 
\begin{equation}
P_{t_k}\left(t_k | s \right) = \delta\left(t_k-\taup k - s\right).
\end{equation}
Since each arrival is deterministic, the joint distribution of all arrivals with known starting point is the product of the marginal distributions,
\begin{equation}
P_{t_1, \dots, t_K}\left(t_1, \dots, t_K | s \right) = \prod\limits_{k=1}^K \delta\left(t_k-\taup k - s\right).
\end{equation}
To account for the fact that the periodicity but not the actual arrival times is known, the starting point is randomly and uniformly chosen in the interval $[0, \taup]$,
\begin{equation}
P_s(s) = \begin{cases} \taup^{-1}, & 0<s<\taup \\ 0, & \text{else} \end{cases} 
\end{equation}

The Fourier transform of the forcing is
\begin{equation} \label{eq:fourier-f0}
\fourT{F_K\left(\frac{t}{\taud}\right)}{\omega}=\frac{\taud}{\sqrt{T}} \sum_{k=1}^K A_k \exp(-i \omega t_k).
\end{equation}
Multiplying this expression with its complex conjugate and averaging over all random variables gives after some calculations the frequency power spectrum of the forcing,
\begin{equation} \label{eq:spectrum-periodic}
\lim_{T \to \infty} \m{\abs{\fourT{F_K}{\omega}^2}} = \frac{\taud^2}{\taup}\,\Arms^2 + \frac{\taud^2}{\taup}\,\Aave^2 2\pi\sum_{n=-\infty}^{\infty} \delta(\taup\omega-2\pi n) .
\end{equation}
According to \Eqref{eq:psd_sn_start}, this is to be multiplied by the spectrum of the pulse function. Thus, the frequency power spectral density for a super-position of periodic pulses with fixed shape and duration is given by the sum of the spectrum of the pulse function (due to a random distribution of pulse amplitudes and represented by the term proportional to $\Arms^2$ in the above equation) and the spectrum of the pulse function multiplied by a uniform delta pulse train, also known as a Dirac comb (represented by the last term in the above equation proportional to $\Aave^2$, which vanishes for a symmetric amplitude distribution).

\end{document}